\documentclass{aa}
\usepackage{graphicx}
\usepackage{txfonts}
\usepackage{natbib}
\usepackage{ifthen}
\title{Monte-Carlo radiative transfer simulation of the circumstellar disk of the Herbig Ae star HD~144432
  \thanks{
    Based on observations made with ESO telescopes at Paranal Observatory under
    program ID 083.D-0224(C) and 085.C-0126(A).
  }
}

\titlerunning{MCRT simulation of the circumstellar disk of HD~144432}

\author{L. Chen\inst{1}
\and A. Kreplin\inst{1}
\and G. Weigelt\inst{1}
\and K.-H. Hofmann\inst{1}
\and D. Schertl\inst{1}
\and F. Malbet\inst{2} 
\and F. Massi\inst{3} 
\and R. Petrov\inst{4} 
\and Ph. Stee\inst{4} 
}

\institute{
Max-Planck-Institut f\"ur Radioastronomie, Auf dem H\"{u}gel 69, 53121 Bonn, Germany
\\email: lchen@mpifr-bonn.mpg.de
\and UJF-Grenoble 1/CNRS-INSU, Institut de Plan\'etologie et d'Astrophysique de Grenoble (IPAG) UMR 5274,
	Grenoble, F-38041, France
\and INAF -- Osservatorio Astrofisico di Arcetri, Largo E. Fermi 5, 50125 Firenze, Italy
\and Laboratoire Lagrange, UMR7293, Universit\'e de Nice Sophia-Antipolis,  
CNRS, Observatoire de la C\^ote d'Azur, 06300 Nice, France
}

\newcommand{\TGM}{temperature-gradient model}

\newcommand{\RT}{radiative transfer}
\newcommand{\MCRT}{Monte-Carlo radiative transfer}

\newcommand{\CP}{closure phase}
\newcommand{\PA}{position angle}
\newcommand{\BD}{brightness distribution}
\newcommand{\SED}{spectral energy distribution}
\newcommand{\enDash}{{\textrm{--}}}
\newcommand{\IR}{infrared}
\newcommand{\NIR}{near-infrared}
\newcommand{\MIR}{mid-infrared}
\newcommand{\Kelvin}{\mathrm{K}}
\renewcommand{\sun}{\odot}
\newcommand{\degree}{\degr} 

\newcommand{\AU}{{\mathrm{AU}}}
\newcommand{\pc}{{\mathrm{pc}}}
\newcommand{\mum}{{\mu\mathrm{m}}}
\newcommand{\pSigma}{{p_\Sigma}}
\newcommand{\DiskTau}{{\tau(2~\mum)}}
\newcommand{\TauMidPlane}{\DiskTau}

\begin{document}

\newlength{\ColumnWidth}
\setlength{\ColumnWidth}{ 9.0cm}

\abstract
{ 
  {
    Studies of pre-transitional disks,
            with a gap region between the inner \NIR-emitting region and the outer disk,
            are important to improving our understanding of disk evolution and planet formation.
  }
    Previous {\IR} interferometric observations have
    shown hints of a gap region in the protoplanetary disk
    around the Herbig Ae star HD~144432.
}
{ 
  We study the dust distribution around this star with two-dimensional
  {\RT} modeling.
}
{ 
  We compare the model predictions obtained via the {\MCRT} code RADMC-3D
    with {\IR} interferometric observations and the {\SED} of HD~144432.
}
{ 
  The best-fit model that we found 
    consists of
    an inner optically thin component at $0.21\enDash0.32~\AU$
    and an optically thick outer disk at $1.4\enDash10~\AU$.
    We also found an alternative model in which the inner sub-AU region
    consists of an optically thin and an optically thick component.
}
{ 
  Our modeling suggests an optically thin component exists
    in the inner sub-AU region,
    although an optically thick component
    may coexist in the same region.
    Our modeling also
    suggests a gap-like discontinuity in the disk of HD~144432.
}


\maketitle

\section{Introduction}
\iftrue

\begin{table*}
\caption{Observation log of our VLTI/AMBER observation of HD~144432.}
\label{obslog}
\centering
\begin{tabular}{cccccccc}
\hline
\hline
Data set  & Night  & $t_{\rm obs}$ & Telescope configuration & $B_{\rm p}$ & PA   & Seeing & DIT\tablefootmark{a}\\ 
 &             & (UTC)         &      & (m)         & ($^\circ$) & ($\arcsec$) & (ms) \\
\hline
2009   & 2009-04-18 & 04:02:36 & E0-G0-H0 & 13.4 / 26.7 / 40.1 & 41.7                & 0.72 & 200   \\
2010a  & 2010-04-18 & 07:57:31 & D0-H0-G1 & 62.5 / 71.2 / 71.3 & 76.9 / 140.6 / 12.8 & 0.75 & 200 \\
2010b  & 2010-04-18 & 10:20:02 & D0-H0-G1 & 46.2 / 67.0 / 68.9 & 92.4 / 164.8 / 24.5 & 0.59 & 200 \\
  \hline
\end{tabular}
  \flushleft \tablefoottext{a}{Detector integration time.}
\end{table*}

In recent years, a growing number of protoplanetary disks have been found
to have a gap region between the inner \NIR-emitting region and the outer disk
\citep[e.g.,][]{2008ApJ...682L.125E%
  ,2010A&A...511A..75B%
  ,2011A&A...531A..93M%
  ,2012ApJ...752..143H%
  ,2013A&A...555A..64M%
  ,2014A&A...561A..26M}.
The mechanisms possibly involved in the formation of a gap include
dynamical clearing by a giant planet,
grain growth, and photoevaporation \citep{2011ARA&A..49...67W}.
These pre-transitional disks are likely in the intermediate evolutionary stage
between an early-stage disk that is optically thick from the dust sublimation radius to the outer edge
and a transitional disk, whose inner region is already cleared.
Studies of pre-transitional disks are important to improve our understanding of
disk evolution and planet formation.

\object{HD~144432} is an isolated Herbig Ae star with spectral type A9/F0,
and is likely a member of the star association SCO-B 2-2 at a distance of $145\pm20~\pc$
\citep[][see also the discussion therein about the distance]{2004A&A...416..647P}.
The strong {\NIR} (NIR) excess of this Herbig Ae star
is a clear signature of hot dust.

In our previous simple {\TGM} of HD~144432 derived from \IR{} (IR) interferometric observations \citep{2012A&A...541A.104C},
the disk consists of two components.
The inner region of the model disk is a narrow ring with an inner radius of ${\sim}0.21~\AU$,
and an inner temperature of ${\sim}1600~\mathrm{K}$.
The outer disk region extends from ${\sim}1~\AU$ to ${\sim}10~\AU$
with an inner-edge temperature of ${\sim}400~\mathrm{K}$.
The lack of IR emission in the region from ${\sim}0.23~\AU$ to ${\sim}1~\AU$ is
possibly a signature of a discontinuity in the dust distribution,
and therefore a signature of the pre-transitional nature of the disk.
However, there is an alternative interpretation  of the emission-free region,
which is that dust exists in this region but is shadowed by a puffed-up rim \citep{2001ApJ...560..957D}
so that it is too cold to emit effectively at IR wavelengths.
In our previous work,
we also found that an extended component that scatters stellar light is needed
to account for the fact that even at short interferometric baselines (${\sim}9\enDash20~\mathrm{m}$),
the NIR visibilities are lower than one.
Plausible origins of this scattering material include an infalling remnant envelope,
dust entrained in the stellar wind/outflow \citep{2006ApJ...647..444M},
or a flaring outer disk \citep{2008ApJ...673L..63P}.
To obtain a more detailed modeling
than in our previous paper
of all available interferometric data,
we performed radiative transfer modeling
and present this in the present paper.

In this paper, we compare two-dimensional {\RT} (RT) models
with all available interferometric observations of HD~144432
to study the dust distribution around the object.
We describe the data set in Sect. 2.
The modeling process and model results are presented in Sect. 3.
We discuss the results in Sect. 4 and present the conclusions in Sect. 5.

\fi

\iftrue

\section{Observations and data reduction}
We observed HD~144432 on 2009 Apr 18 and 2010 Apr 18 with the VLTI/AMBER instrument (see Table 1).
The AMBER instrument is a near-infrared beam combiner
that records three-beam interferograms \citep{Petrov2007}.
The instrument was operated in the low spectral resolution mode (R=35) and recorded fringes in the $H$ and $K$ bands simultaneously.
The fringe tracker FINITO was not used.
The observations of HD~144432 were conducted using the linear baseline configuration E0-G0-H0%
\footnote{
  \tiny
    http://www.eso.org/sci/facilities/paranal/telescopes/vlti/configuration
}
and the triangle configuration D0-H0-G1
of the 1.8~m auxiliary telescopes (ATs).
For data processing, we used the AMBER data reduction package amdlib 3.0%
\footnote{
http://www.jmmc.fr/data\_processing\_amber.htm
}.
We selected 30\% of the target and calibrator frames with the highest fringe signal-to-noise-ratio (SNR)
  to improve the visibility SNR \citep{Tatulli2007} of each target and calibrator data set.
  Furthermore, we applied a method that equalizes the histograms of the optical path differences
  of the target and calibrator interferograms \citep{2012A&A...537A.103K}
  to improve the calibration of the visibilities.
  On both nights we used the calibrator star HD~142669,
  which has a uniform disk diameter of $d_{\mathrm{UD}} = 0.27 \pm 0.05$ mas%
  \footnote{
    Taken from the Catalogue of Stellar Diameters \citep[CADARS;][]{Pasinetti-Fracassini2001}
  }.

We complemented our data with existing data from the literature,
including the $K$-band observation with KI \citep{2005ApJ...624..832M,2009ApJ...692..309E},
the $H$-band observation with IOTA \citep{2006ApJ...647..444M},
the {\MIR} (MIR) interferometry with VLTI/MIDI \citep{2004A&A...423..537L},
as well as the {\SED} (SED) \citep{2004A&A...423..537L}.

\fi

\section{Modeling}
\iftrue

\newcommand{\TabularParameterDescription}{PARAM: Undefined!}
\newcommand{\TabularContent}{Tabular: undefined!}
\newcommand\StandardTableAndFigure{}
\newcommand{\FigRef}{}
\newcommand{\StandardTableAndFigureList}{}

\newcommand{\ModelFigureCaptionTextOne}{
  Model scanning run {\RunNameShort} (\ModelClassDescription),
  and the model parameters of the best model {\ModelNameShort}
  with minimum $\chi^2_\mathrm{\DatasetName,red}$,
  which is calculated by comparing each model with data set \DatasetName.

  {\it a:}
  Parameter ranges of model scanning run {\RunNameShort},
  number of tested parameter values (steps) per parameter,
  and parameters of the best model {\ModelNameShort}.
  {
    \TabularParameterDescription
  }

  {\it b:}
  $\chi^2_\mathrm{\DatasetName,red}$ maps of Model run \RunName.
  {
    For each subset of parameters, the $\chi^2_\mathrm{\DatasetName,red}$
    shown is the lowest value for all combinations with other parameters.
    For example, for each pair of ($R_\mathrm{in}$, $R_\mathrm{out}$) values,
    the $\chi^2_\mathrm{\DatasetName,red}$ values for all possible
    ($h_\mathrm{in}$, $p_{h}$) combinations within the described ranges
    were compared
    and the minimum value found
    is plotted into the map at left panel.
  }

  {\it c:} 
  {
    The dust density distribution of the model,
    in units of $\mathrm{g}~\mathrm{cm}^{-3}$.
  }

  {\it d:}
  {
    $K$- (left) and $N$-band (right) model images A1,
    in units of $\mathrm{erg}~\mathrm{cm}^{-2}~\mathrm{s}^{-1}~\mathrm{Hz}^{-1}$.
  }

  {\it e:} NIR and MIR visibilities, and closure phases CP
  (red dots: observations; black lines: model).
  The VLTI data are our own data.
  The data sets 2004 IOTA, 2003 KI, 2007 KI, and 2003 MIDI mentioned in the panels
  are taken from
  \citet{2006ApJ...647..444M}, \citet{2005ApJ...624..832M},
  \citet{2009ApJ...692..309E}, and \citet{2004A&A...423..537L},
  respectively.

  {\it f:} SED \citep[from][]{2004A&A...423..537L}.
  The lines denote the model contributions from different radial regions.
  The red dots are the observations.

}
\newcommand{\ModelFigureCaptionTextTwo}{
  Same as Fig.~1, except for model run {\RunName} and compared with data set \DatasetName.
  Model {\ModelNameShort} is the model with minimum $\chi^2_\mathrm{\DatasetName,red}$.
  {
    \TabularParameterDescription
  }
}

\newcounter{CounterModel}
\newboolean{FirstModel}
\newcommand{\RunNameShort}{}
\newcommand{\RunNameLong}{}
\newcommand{\RunName}{\RunNameShort}
\newcommand{\DatasetName}{}
\newcommand{\ModelNameLong}{}
\newcommand{\ModelNameShort}{\RunName\DatasetName}
\newcommand{\ModelClassDescription}{}
\newcommand{\ModelName}{\ModelNameLong}
\newcommand{\ModelFigureCaptionText}{}
\newcommand{\TableName}{}
\newcommand{\BestChiSquare}{$some value??$}

  \setlength\fboxsep{0pt}
  \setlength\fboxrule{0.1pt}
\newcommand{\SubFig}[2]{
  \mbox{
    \parbox[t]{0cm}{
      \vspace{0.1cm}
      \mbox{\large \bf
        {#1}
      }
    }
    \nolinebreak
    \parbox[t]{1.0\ColumnWidth}{%
      \vspace{0mm}
      \parbox[b]{1.0\ColumnWidth}{#2}
    }
  }
}

%

\renewcommand\StandardTableAndFigure{

  \stepcounter{CounterModel}
  \ifthenelse{\arabic{CounterModel}=1}{\setboolean{FirstModel}{true}}%
  {\setboolean{FirstModel}{false}}

  \renewcommand{\TableName}{table:parameters:\RunNameLong_dataset\DatasetName}
  \renewcommand{\ModelName}{MCRT_\RunNameLong_dataset\DatasetName}

  
\renewcommand{\TabularContent}{ 
  \begin{tabular}{ccccc}
    \hline
    \hline
    \HeadLineForModelTabular
    \\
    \hline
    Inner disk
    \\
    $R_\mathrm{in}$ $[\AU]$
    & $0.196$
    & $-$
    & $0.196$
    \\
    $W_\mathrm{ring}$
    & $0.5$
    & $-$
    & $0.5$
    \\
    $\pSigma$
    & $0$
    & $-$
    & $0 $
    \\
    $h_\mathrm{in}$
    & $0.05\enDash1.0$
    & $21$
    & $0.35$
    \\
    $\tau$
    & $0.1 \enDash3.0$
    & 21
    & $0.54$
    \\
    $f_\mathrm{carbon}$
    & $0.5$
    & $-$
    & $0.5$
    \\
    $f_\mathrm{small}$
    & $1.0$
    & $-$
    & $1.0$
    \\
    \hline
    $\chi^2_\mathrm{2,red}$
    &
    &
    & \BestChiSquare
    \\
    \hline
  \end{tabular}
}

  \ifthenelse{\boolean{FirstModel}}{
    \renewcommand{\ModelFigureCaptionText}{\ModelFigureCaptionTextOne}
  }{
    \renewcommand{\ModelFigureCaptionText}{\ModelFigureCaptionTextTwo}
  }

  \begin{figure*}
    \begin{minipage}[t]{2.0\ColumnWidth}
      \begin{minipage}[t]{\ColumnWidth}
        \SubFig{a}{
          \parbox{\ColumnWidth}{ 
            \setlength{\tabcolsep}{9pt}
            \renewcommand{\arraystretch}{1.2}
            \hspace{0.5cm}
            \parbox{0.8\ColumnWidth}{
              \setlength{\baselineskip}{10pt}
              \TabularContent
            }
          }
        }

        \SubFig{b}{
          \includegraphics[width=1.0\ColumnWidth,angle=-0]{\ModelName.chi2.ps}
        }

        \SubFig{c}{
          \includegraphics[scale=0.35,angle=-90,,trim=2cm 2cm 1.8cm 0cm]{%
            \ModelName.dust_density_logxz.ps%
          }
        }

        \SubFig{d}{
          \hfill
          \mbox{\includegraphics[scale=0.15,angle=-0]{\ModelName.image_K.ps}}
          \hspace{0.0cm}
          \nolinebreak
          \mbox{\includegraphics[scale=0.15,angle=-0]{\ModelName.image_N.ps}}
          \hfill
          \rule{0mm}{0mm}
        }

      \end{minipage}
      \begin{minipage}[t]{\ColumnWidth}

        \SubFig{e}{
          \parbox{\ColumnWidth}{
            \includegraphics[scale=0.25,angle=-0]{\ModelName.V_NIR.eps}
            \linebreak
            {
              \mbox{\includegraphics[scale=0.25,angle=-0]{\ModelName.V_MIR.eps}}%
              \nolinebreak%
              \mbox{\includegraphics[scale=0.25,angle=0]{\ModelName.CP.eps}}
            }
          }
        }

        \SubFig{f}{
          \mbox{\includegraphics[scale=0.35,angle=-90,trim=0 0 -0.2cm 0]{\ModelName.SED.ps}}
        }

      \end{minipage}
    \end{minipage}
    \caption{ \ModelFigureCaptionText }
    \label{figure:\ModelName}
  \end{figure*}

}

\renewcommand{\FigRef}{%
  \ref{figure:MCRT_\RunNameLong_dataset\DatasetName}%
}
\renewcommand{\StandardTableAndFigureList}{%
  Fig.~\FigRef%
}

\newcommand{\HeadLineForModelTabular}{%
  Parameter                 & Parameter range                   & Steps     & Model \ModelNameShort
}

To interpret all the interferometric data,
we employ {\MCRT} (MCRT) modeling
using the code RADMC-3D%
\footnote{
http://www.ita.uni-heidelberg.de/~dullemond/software/radmc-3d/},
which is the follow-up version of RADMC \citep{2004A&A...417..159D}.
The code RADMC has been intensively used in modeling circumstellar disks
\citep[e.g.,][]{2003A&A...398..607D,2007ApJ...656..980P,2013A&A...551A..21K}.
Both RADMC and RADMC-3D use the instant-re-emission method \citep{1999A&A...344..282L}
and the immediate temperature correction method \citep{2001ApJ...554..615B}.

In the modeling process, we assume a distance of $145~\mathrm{pc}$
and a stellar luminosity of $7.6~L_\sun$
\citep{2004A&A...416..647P}.
For the star {SED}, we adopt the same Kurucz model as in our previous paper \citep{2012A&A...541A.104C}.
The dust consists of astronomical silicate \citep{1984ApJ...285...89D}
and amorphous carbon \citep{1998A&A...332..291J}.
We assume a grain size distribution with $n(a)\propto a^{-3.5}$
with small grains ($a_\mathrm{min}=0.01~\mum$, $a_\mathrm{max}=1.0~\mum$)
and large grains ($a_\mathrm{min}=1.0~\mum$, $a_\mathrm{max}=1000~\mum$).
We use
the parameters $f_\mathrm{small}$ and $f_\mathrm{carbon}$ to control the dust composition,
so that the fractions of mass in small carbonaceous grains,
small silicate grains,
large carbonaceous grains,
and
large silicate grains are
\begin{displaymath}
  f_\mathrm{small,carbon} =
  f_\mathrm{small} f_\mathrm{carbon}
  ,
\end{displaymath}
\begin{displaymath}
  f_\mathrm{small,silicate} =
  f_\mathrm{small} (1-f_\mathrm{carbon})
  ,
\end{displaymath}
\begin{displaymath}
  f_\mathrm{large,carbon} =
  (1-f_\mathrm{small}) f_\mathrm{carbon}
  ,
\end{displaymath}
and
\begin{displaymath}
  f_\mathrm{large,silicate} =
  (1-f_\mathrm{small}) (1-f_\mathrm{carbon})
  ,
\end{displaymath}
respectively.
We assume thermal coupling between the disk species.
In the calculation of each visibility,
we take into account the field-of-view (FOV) effect of the interferometer used
by multiplying the predicted {\BD} with a mask
with a diameter of the interferometer FOV
(VLTI/AMBER: 0.25\arcsec;
KI: 0.05\arcsec;
IOTA/IONIC3: 0.8\arcsec;
VLTI/MIDI: 0.52\arcsec%
).

Our previous {\TGM}ing \citep{2012A&A...541A.104C}
suggests that the disk is oriented roughly face-on,
with the inclination of ${\sim}10\degree$
and a {\PA} (PA) of the major disk axis of ${\sim}30\degree$.
In the following,
we adopt this inclination angle and PA %
of our previous best-fit {\TGM}.

Our disk models and results are presented below.
We first try to find models that can roughly reproduce
the visibilities and the SED at $\lambda<15~\mum$.
This subset of data (hereafter ``data set 1") is chosen
because it is sensitive to
a relatively small parameter set
and the computational cost
of the parameter-space scanning can therefore be reduced.
We only further refine the model to fit the full data set
(hereafter ``data set 0"),
including the SED data at $\lambda>15~\mum$,
when the fitting of data set 1 is sucessful.
  \newcommand{\ChiSquareDef}[1]{
    \sum\limits_{i=1}^{N_{{#1}}} \left(
      \frac{
      {#1}_{i,\mathrm{model}} - {#1}_{i,\mathrm{obs}}
    }{
      \Delta{#1}_{i,\mathrm{obs}}
    }
      \right) ^2
  }
  To quantitatively measure the deviation of
  the model predictions
  from a given data set,
  we define
  \begin{displaymath}
    \chi^2_V = \ChiSquareDef{V}
    ,
  \end{displaymath}
  \begin{displaymath}
    \chi^2_{CP} = \ChiSquareDef{CP}
    ,
  \end{displaymath}
  \begin{displaymath}
    \chi^2_\mathrm{SED} = \ChiSquareDef{f}
    ,
  \end{displaymath}
  and
  \begin{displaymath}
    \chi^2_{\mathrm{red}}
    = \frac
    { 
      \chi^2_V + \chi^2_{CP} + \chi^2_\mathrm{SED}
    }
    { 
      N_V + N_{CP} + N_{f}
      - N_{p}
    }
    ,
  \end{displaymath}%
  in which $V$, $CP$, and $f$ denote visibilty, closure phase, and flux, respectively,
  and $N_V$, $N_{CP}$, $N_{f}$ are the numbers of 
  observational points shown in the figures.
  The parameter $N_{p}$ is the number of free parameters in a model setting.

{
  We use spherical coordinates $(r,~\theta,~\varphi)$ throughout the paper.
}

\renewcommand{\RunNameLong}{1comp}
\renewcommand{\RunNameShort}{A}
\renewcommand{\DatasetName}{1}
\renewcommand{\ModelClassDescription}{simple one-component disks}
\renewcommand{\ModelFigureCaptionText}{\ModelFigureCaptionTextOne}
\renewcommand{\BestChiSquare}{$8.56$}
\subsection{Simple power-law disk with vertical rim (Model {\ModelNameShort})}
\StandardTableAndFigure

We start with a simple disk model in which the surface density of the disk
obeys the power-law distribution
\begin{math}
\Sigma= \Sigma_{\mathrm{in}}
\left(
{R}/{R_\mathrm{in}}
\right)^{\pSigma}
\end{math}
for 
\begin{math}
R_{\mathrm{in}}\leq R \leq R_{\mathrm{out}}
\end{math}.
{
  For the vertical distribution of the gas, we assume the Gaussian function
  $\rho_\mathrm{gas}(R,\theta)
  =
  \Sigma(R) \frac{1}{\sqrt{2\pi}H} \exp\left[ -\frac{R^2(\frac\pi2-\theta)^2}{2H^2} \right]$.
}
For the scale height $H$, we assume the power-law distribution
$
h\equiv \frac{H}{R} = h_{\mathrm{in}}
\left(
  {R}/{R_\mathrm{in}}
\right)
^{p_h}
$.
Therefore, the free parameters involved here are the inner and outer radii $R_\mathrm{in}$ and $R_\mathrm{out}$,
the power-law indices $p_\Sigma$ and $p_h$,
the dimensionless scale height $h_\mathrm{in}$ at the inner radius,
  the parameters $f_\mathrm{small}$ and $f_\mathrm{carbon}$ to control the dust composition,
and the mass scaling factor $\Sigma_\mathrm{in}$,
or equivalently, the total dust mass, or the mid-plane optical depth $\TauMidPlane$.
The mid-plane optical depth is calculated as
\begin{displaymath}
  \TauMidPlane =
  \kappa
  \int_{R_\mathrm{in}}^{R_\mathrm{out}}
  \rho_\mathrm{gas}(R,0)
  d R
  ,
\end{displaymath}
where $\kappa$ is the absorption coefficient of the dust at $2~\mum$.
We tested a large number of RT models (see Fig.~{\FigRef}a)
and calculated the reduced chi-square $\chi^2_\mathrm{1,red}$ for each model
by comparing the model predictions with data set 1.
We find that $\chi^2_\mathrm{1,red}$ mainly depends on 
the four parameters $R_\mathrm{in}$, $R_\mathrm{out}$, $h_\mathrm{in}$,
and $p_h$,
while it shows only weak dependence on other parameters
(for $\TauMidPlane\gg1$; see Appendix A for more details).
Therefore, in our model scanning run A, we computed models
on a grid of these four parameters,
with five to ten different parameter values (steps) per parameter
and a total model number of $7\times5\times10\times9=3150$.
The ranges of the scanned parameter and the best model 
are presented in {\StandardTableAndFigureList}.
None of these models is able to approximately reproduce data set 1.
  For the best model {\ModelNameShort}
  (see table in Fig.~{\FigRef}a),
  the $\chi^2_\mathrm{1,red}$ is {\BestChiSquare},
  and the individual contributions to $\chi^2$ from each data subset
  are 
  $\chi^2_V=1208.6$,
  $\chi^2_{CP}=14.0$,
  $\chi^2_\mathrm{SED}=711.5$,
  respectively.

\renewcommand{\RunNameLong}{1comp_curved}
\renewcommand{\RunNameShort}{B}
\renewcommand{\DatasetName}{1}
\renewcommand{\ModelClassDescription}{one-component disks with curved rims}
\renewcommand{\BestChiSquare}{$8.94$}
\subsection{Disk with curved inner disk rim (Model {\ModelNameShort})}
\StandardTableAndFigure

Because of the density dependency of the sublimation temperature and the dust settling effect,
the inner disk rim is predicted to be curved \citep{2005A&A...438..899I,2007ApJ...661..374T}.
Therefore, we also tested models with a curved rim
to investigate whether the rim curvature improves the model fitting.
We add the curvature to the rim in a parameterized way.
The inner radius of the disk is assumed to increase with height
$
R_\mathrm{in} (\theta) = R_\mathrm{in} (0) \left[
  1+f_\mathrm{curv}\left(\frac\pi2-\theta\right)^2
\right]
,$
where $f_\mathrm{curv}$ is a free parameter to control how curved the rim is.
The curvature is implemented as a cut through the density profile
prior to the RT calculations.
Near the mid-plane and $R_\mathrm{in}$, our parameterized
curvature rim has roughly a parabolic profile,
which is similar to those
found in \citet{2005A&A...438..899I}.
For instance,
when the parameter $f_\mathrm{curv}$ is set to 10,
the inner radius at $z/R\approx \frac\pi2-\theta =0.1$
is $10\%$ larger than $R_\mathrm{in}(0)$.
This curvature is comparable with the self-consistent solution in
(\citealt{2005A&A...438..899I}; see Fig.~2 therein).

In our model scanning run B, we computed models
on a grid of five parameters,
with a total model number of $5\times3\times6\times7\times4=2520$.
The scanned parameter ranges and the best model 
are presented in \StandardTableAndFigureList.
None of these models is able to approximately reproduce data set 1.
The best model {\ModelNameShort} has a $\chi^2_\mathrm{1,red}$ of {\BestChiSquare}.

\renewcommand{\RunNameLong}{1comp_shadowed}
\renewcommand{\RunNameShort}{C}
\renewcommand{\DatasetName}{1}
\renewcommand{\ModelClassDescription}{one-component disks with shadowed regions}
\renewcommand{\BestChiSquare}{$7.23$}
\subsection{Self-shadowed Disk (Model \ModelNameShort)}
\StandardTableAndFigure

We decided to test disk models
with a shadowed region behind an optically thick puffed-up inner rim
because single-component disk models were unable to reproduce data set 1.
We set the scale height of such a disk as
\begin{equation}
  h=\left\{
    \begin{array}{ll}
      h_\mathrm{in}+h_\mathrm{puff}
      , & \textrm{for}~ R_{\mathrm{in}}\leq R \leq R_{\mathrm{rim,out}}
      \\
      h_\mathrm{in}
      \left(
        {R} \over {R_\mathrm{in}}
      \right)
      ^{p_h}
      , & \textrm{for}~ R_{\mathrm{rim,out}}\leq R \leq R_{\mathrm{out}}
    \end{array}
    \right.
    ,
  \end{equation}
  where $R_{\mathrm{rim,out}}=R_{\mathrm{in}}(1+W_\mathrm{rim})$.
  Therefore, the additional parameters of this model are the relative width $W_\mathrm{rim}$
  and the additional scale height $h_\mathrm{puff}$ of the puffed-up rim.

In our model scanning run C, we computed models
on a grid of five parameters
with a total model number of $5\times3\times5\times10\times7=5250$.
The scanned parameter ranges and the best model 
are presented in \StandardTableAndFigureList.
None of these models is able to approximately reproduce data set 1.
The best model {\ModelNameShort} has a $\chi^2_\mathrm{1,red}$ of {\BestChiSquare}.

\renewcommand{\RunNameLong}{2comp_ThinThick}
\renewcommand{\RunNameShort}{D}
\renewcommand{\DatasetName}{1}
\renewcommand{\ModelClassDescription}{two-component disks}
\renewcommand{\BestChiSquare}{$2.55$}
\subsection{Two-component disk Model {\ModelNameShort}}
\StandardTableAndFigure

The failure of all our disk models consisting of 
a homogeneous, optically thick disk (models A1 and B1)
or a disk consisting of a narrow inner rim, a shadowed region,
and an outer disk region (all of these components are optically thick; model C1),
suggests that
another disk structure is needed to reproduce the data.
In particular,
all these models with an optically-thick inner rim
are unable to reproduce the NIR SED and visibilities
(see also the additional studies in Appendix B).
On the other hand,
we found that with an optically thin dust
  ($\tau<1$ at $2~\mum$)
located in the sub-AU region,
it is possible to roughly reproduce the NIR observations (Appendix B.4).
Therefore, we additionally tested models with
an inner optically thin component and an outer optically thick component.
In our model scanning run D, we computed models
on a grid of five parameters,
with a total model number of $5\times5\times5\times5\times5=3125$.
The parameter ranges and results (model {\ModelNameShort})
are shown in \StandardTableAndFigureList.
We found that, with this class of models, the $\chi^2_\mathrm{1,red}$ can be reduced to {\BestChiSquare},
much lower than in any of the previous models.
In model {\ModelNameShort}, the computed NIR excess consists of
thermal emission from the inner disk,
scattered stellar light from the inner disk,
and scattered light from the outer disk.
The mass of the optically thin inner disk is $1.3\times10^{-12}M_\sun$.

\renewcommand{\RunNameLong}{2comp_ThinThick_SubMM}
\renewcommand{\RunNameShort}{DA}
\renewcommand{\DatasetName}{0}
\renewcommand{\ModelClassDescription}{two-component disks}
\renewcommand{\BestChiSquare}{$2.50$}
\StandardTableAndFigure

\subsection{Two-component disk Model \ModelNameShort}
A problem of the models in run D is that, with the carbon/silicate ratio of 1:1,
they cannot reproduce the strong $10~\mum$ silicate feature in the SED.
Therefore, we also tested models with different carbon/silicate ratio.
In our model scanning run \RunNameShort,
we set carbon/silicate=1:9.
We also added large grains into the model
so that the SED data at $\lambda>15\mum$ can also be reproduced.
The parameter ranges and results 
are shown in \StandardTableAndFigureList.
The model {\ModelNameShort} approximately reproduced the whole data set 0,
including the silicate feature and the SED data at $\lambda>15\mum$.
In model {\ModelNameShort}, similar to model D1,
the computed NIR excess consists of
thermal emission from the inner disk,
scattered stellar light from the inner disk,
and scattered light from the outer disk.
The mass of the optically thin inner disk is $1.8\times10^{-11}M_\sun$.

\renewcommand{\RunNameLong}{3comp_DoubleThick_SubMM}
\renewcommand{\RunNameShort}{E}
\renewcommand{\DatasetName}{0}
\renewcommand{\ModelClassDescription}{optically thin and thick dust at sub-AU scale}
\renewcommand{\BestChiSquare}{$2.93$}
\StandardTableAndFigure

\subsection{Three-component disk Model \ModelNameShort}
The purpose of parameter scanning run {\RunNameShort}
is to test the possibility of modeling the data
with a two-component inner disk
consisting of both an optically thin and an optically thick component
in addition to
an optically thick outer disk.
The goal of this study is to investigate whether
we can really discriminate in our study between
(i) an inner optically thin disk
and
(ii) an inner optically thin disk with an additional optically thick core.
The parameter ranges and results 
are shown in \StandardTableAndFigureList.
The model {\ModelNameShort} roughly reproduced data set 0,
with a $\chi^2_\mathrm{0,red}$ (\BestChiSquare) only slightly higher than that of model DA0 (2.50).
This small $\chi^2$ difference is probably too small to
reliably discriminate between these two models.

\fi

\clearpage
\clearpage

\section{Discussion}
\iftrue
\subsection{Optical depth and composition of the inner NIR-emitting dust}

In our modeling we found that an optically thin dust located at the sub-AU region
is needed to reproduce the NIR SED and visibilities.
In our previous {\TGM}ing \citep{2012A&A...541A.104C},
a blackbody emission with $T{\sim}1600~\Kelvin$ at ${\sim}0.2~\AU$ from the central star was needed
to reproduce the NIR SED and visibilities.
Our {RT} modeling presented here suggests that this kind of hot emission at this location
requires
an optically thin dust composed of small grains.
All the models tested with an optically thick rim at ${\sim}0.2~\AU$,
on the other hand,
predict a temperature that is too low,
and therefore, a red SED.
In other words, the SEDs they predict are flatter in $2\enDash7~\mum$ than the observed SED.
Therefore, they either underestimate the flux at ${\sim}2~\mum$ (as in model A1, B1, and C1),
or overestimate both the flux at ${\sim}7~\mum$ and the MIR visibilities
(as in model A2, B2, and C2 shown in Appendix B).

The temperature of a passively heated dust depends on both
its optical depth and the wavelength dependency of its optical properties
\citep[e.g.,][]{2010ARA&A..48..205D}.
If an optically thin gray dust lies at $0.2~\AU$ from a central heating source of $L_*=7.6~L_\sun$,
its radiative-equilibrium temperature would be
$T_\mathrm{gray}=
\left(
\frac{L_*}{16\pi\sigma{R^2}}
\right)^{1/4}
{\sim}1000\Kelvin$.
However, astronomical dust is non-gray,
typically with an opacity lower in the IR than in the UV/optical,
{
  and therefore with a cooling parameter $\epsilon<1$
    \citep[for definition of this parameter see Eq. 10 in][]{2010ARA&A..48..205D}.
}
This kind of non-gray dust will be overheated to
$T=T_\mathrm{gray}\left(\frac1\epsilon\right)^{\frac14}$
\citep[see, e.g., Eq. 9 in][]{2010ARA&A..48..205D}.
Therefore, by choosing dust with $\epsilon\ll1$ (corresponding to small grain size),
  it is possible to reproduce the observed blue NIR emission.

The situation for optically thick dust is different.
Within optically thick dust with $\epsilon\ll1$,
only a thin surface layer is subject to the
overheating effect.
The inner parts are much cooler
\citep[see, e.g., Eq. 11 and Fig. 9 in][]{2010ARA&A..48..205D}.
The overall emission of this kind of dust is a mixture of
the blue emission from the surface layer and the redder emission from the inner body.
Therefore, the overall color of the emission is not substantially bluer than
that from an optically thick gray dust.
This insensitivity of the NIR SED to the $\epsilon$ parameter has been confirmed
by our parameter scanning run AS (see Appendix A),
   which covers a wide range of $\epsilon$
   by choosing different dust compositions.

{
  Therefore, we conclude that the hot NIR emission is likely dominated
    by an optically thin component.
    However, it cannot be excluded that
    an optically thick component may coexist in the sub-AU region (e.g., model E0).
    This component likely has low scale height ($<0.01$),
  so that it intersects only a small fraction of stellar light
  and the NIR emission is still dominated by the optically thin component.
}

\subsection{Overall structure of the CS disk}

We compared several kinds of MCRT models with the observations.
We found that all the models with the NIR emission dominated
by an optically thick inner rim,
without an optically thin component,
are unable to reproduce the observations
(see Sect. 3.1, 3.2 and 3.3 for the model scanning).
The best model that we found (model D0)
consists of
an inner optically thin component
and
an outer optically thick component.
However, we cannot exclude the alternative model 
with an
inner sub-AU region consisting of
both optically thin
and optically thick components (model E0).
The additional optically thick component in this model
has a low scale height
and the NIR emission is still dominated by the optically thin component.

\citet{2001A&A...365..476M%
} classified the Herbig Ae/Be stars into two groups,
  based on the shape of their SED.
  The SEDs of group I objects can be fitted with a power law and a blackbody
  and the SEDs of group II objects can be fitted with merely a power law.
  The different spectral shapes have been interpreted
  in the following way:
  group I objects have flaring disks and group II objects have self-shadowed disks
  \citep{2004A&A...417..159D}.
  In recent years,
  evidence of disk gaps has been found in a growing number of
  group I objects
  \citep[e.g.,][]{
    2010A&A...511A..75B%
      ,2011A&A...531A..93M%
      ,2012ApJ...752..143H%
      ,2013A&A...555A..64M%
      ,2014A&A...561A..26M%
  }.
Therefore, it has been suggested that the gap region is a common feature
of most group I Herbig stars,
   and the blackbody part of a SED of that type
   is created by the illuminated inner wall of the outer disk
   \citep{2012ApJ...752..143H,2013A&A...555A..64M}.
   Recent studies of group II objects with high spatial resolution
   provide evidence that some of them also contain gaps.
   For example, \citet{2013A&A...555A.103S} fitted the observation of the group IIa object HD~142666
   with a {RT} model with a disk gap
   from $0.35~\AU$ to $0.8~\AU$.
   \citet{2015A&A...581A.107M} found that a population of group II disks
   have large MIR half-light radii,
   which are more consistent with gapped disks rather than continuous disks.
{
  Taking advantage of high-resolution {\IR} interferometry
    and {MCRT} modeling,
  we find evidence of a gap region in the group II object HD~144432.
    This result reinforced the aforementioned suggestion by \citet{2015A&A...581A.107M}.
}
The gap in our model of HD~144432 has an outer radius of only ${\sim}1.4~\AU$,
    while the reported gap outer radii of group I objects range from $5.6~\AU$ to $63~\AU$.

    The disk gap in our model is possibly a signature of planet formation in the disk
    \citep{1999ApJ...514..344B%
      ,2011ApJ...729...47Z%
        ,2012ARA&A..50..211K%
        ,2015ApJ...808L...3A%
    }.
Grain growth of dust is another mechanism that is able to create disk cavities.
However, numerical simulations of grain growth \citep{2012A&A...544A..79B}
predict inside-out growth of dust
and smooth radial distribution of dust size%
, which seems to be inconsistent with the dust distributions in our models.
Therefore, grain growth is unlikely to be the only mechanism to explain
the dust distribution found in our modeling.
Furthermore, photoevaporation clearing is another possible mechanism
that is able to create the gap in the disk of HD~144432.
According to numerical simulations of the photoevaporation clearing process
\citep{2006MNRAS.369..229A,2007MNRAS.375..500A,2010MNRAS.401.1415O},
  the first step of disk clearing is the opening of a gap at ${\sim}1~\AU$,
  which prevents the inner disk from being replenished.
  Consequently, the inner disk is quickly drained into the center,
  and then the inner edge of outer disk starts to move outward.
  The gap in HD~144432 has a small outer radius of ${\sim}1.4~\AU$,
  comparable with a newly opened photoevaporative gap.
  Therefore, the disk might be observed in the photoevaporation-starved accretion phase.
{
  This scenario is consistent with the fact that the object is still accreting
    \citep[%
    $\dot{M}\approx8.5\times10^{-8}~M_\sun~\mathrm{yr}^{-1}$,%
    ][]{2006A&A...459..837G}.
}
However, with the current data set, we cannot distinguish between
the various gap-opening mechanisms because, unfortunately,
    the $uv$ coverage of our data set is poor,
    the baseline lengths are too short to fully
    resolve the inner disk region,
    and image reconstruction is not possible with
    this type of $uv$ coverage.
    We also cannot exclude the possibility that more than one mechanism
    is responsible for the disk evolution
    \citep[e.g.,][]{2013MNRAS.430.1392R,2009ApJ...704..989A}.

The inner radii of the inner disk components of the
two best models DA0 (Fig. 5) and E0 (Fig. 6)
are approximately $0.21$ and $0.15~\AU$.
These radii are similar to the dust sublimation radius of $0.13~\AU$
suggested by the size-luminosity relation reported by
\citet{2005ApJ...624..832M} for a dust sublimation temperature of $1500~\Kelvin$
and a luminosity of $7.6~L_\sun$.
To our knowledge, there is no radial velocity measurement
indicating a close binary in this object.
Therefore, it is likely that the inner radius of the inner disk
of HD~144432
is indeed defined by dust sublimation.

For some pre-transitional disks,
    the innermost NIR-emitting regions inside the gaps are suggested to be optically thin 
    \citep{%
      2010A&A...512A..11M,%
        2010A&A...511A..75B%
        ,2010ApJ...717..441E%
        ,2013A&A...555A..64M%
        ,2014A&A...561A..26M%
    }.
The origin of such optically thin dust with large scale height remains unclear.
\citet{2011A&A...531A..80K} proposed that
an optically thin halo can arise from the destructive collision
between planetesimals on highly-inclined orbits.
\citet{2012ApJ...758..100B} proposed that a centrifugally-driven dusty disk wind
can arise from the disk close to the dust sublimation radius.
Our model E0 seems to be compatible with this scenario.

\fi

\section{Summary and conclusions}
\iftrue
We computed two-dimensional {RT} models
for the Herbig Ae star HD~144432.
The following results were obtained.

All computed models with only optically thick components
cannot reproduce the {observations}.
One main problem of such models is that the NIR excess in the SED of this object
cannot be reproduced with an inner optically thick disk rim,
when the constraints on the size of the NIR-emitting region
imposed by the NIR interferometric observations are taken into account.
The best model that we found 
consists of
an inner optically thin component at $0.21\enDash0.32~\AU$
and an optically thick outer disk at $1.4\enDash10~\AU$.
In an alternative model with similar quality, the inner sub-AU region consists of
an optically thin and an optically thick component.
In each of the above models, there exists a gap region,
   and the NIR emission is dominated by the inner optically thin component.
   Therefore, our modeling suggests that the dust distribution
   in the protoplanetary disk of HD~144432
   differs from the dust distribution of an early-stage disk,
   which is optically thick from the dust sublimation radius to the outer edge.

\fi

\iftrue
\begin{acknowledgements}
We thank our ESO colleagues for their excellent support during the observations.
The comments and suggestions from an anonymous referee helped to improve the quality of the paper.
\end{acknowledgements}
\fi

\iftrue

\appendix

\section{Additional parameter study for the simple one-component disk (model type A; Sect. 3.1)}

\renewcommand{\RunNameLong}{1comp_secondary}
\renewcommand{\RunNameShort}{AS}
\renewcommand{\DatasetName}{1}
\renewcommand{\ModelClassDescription}{simple one-component disks}
\renewcommand{\BestChiSquare}{$7.81$}
\StandardTableAndFigure

In our additional model scanning run {\RunNameShort}
presented in \StandardTableAndFigureList,
we kept the
 parameters $R_\mathrm{in}$, $R_\mathrm{out}$, $h_\mathrm{in}$ fixed
and computed models
on a grid of the four parameters
$\pSigma$,
$\TauMidPlane$,
$f_\mathrm{carbon}$,
and $f_\mathrm{small}$,
with a total model number of $5\times6\times6\times5=900$.
The parameter ranges and results 
are shown in \StandardTableAndFigureList.
The results show that $\chi^2_\mathrm{1,red}$ is almost insensitive to the above free parameters%
\footnote{
  Exceptions to such an insensitivity are discussed below.
  1) For the models with a low $\TauMidPlane$ and a steep surface density profile
  (bottom left region in the left $\chi^2$ panel of Fig. A.1b)
  the outer parts of the disk become optically thin,
  so that the fluxes at long wavelengths are reduced.
  However, the NIR SED stays approximately the same.
  2) For the models with only large grains
  ($f_\mathrm{small}=0$, bottom region in the right panel of Fig. A.1b),
  the disk emits more at long wavelengths and less at NIR,
  in comparison with the models with small grains.
  However, the slope of the NIR SED does not change.
  Therefore, in these regions of parameter space,
  the model still predicts a red NIR SED.
  Therefore, the main problem of models with optically thick inner rim,
  that is, too red NIR SED, cannot be solved by altering these four parameters.
}
.
This supports our decision to keep these parameters constant
in most of our parameter-scanning runs.

\section{Comparing scanned models with data set 2}
In the model scanning runs A, B, and C,
we notice that all these models cannot reproduce
the NIR emission.
To investigate this deviation of the models from the observations,
we compare the model results with
data set 2, which is composed of the following data:
\\~~$-$ all the SED data with wavelengths ${<}7\mum$;
\\~~$-$ all NIR visibilities with baseline length ${>}30$m; and
\\~~$-$ all NIR {\CP}s.\\
In the following sections, we want to investigate whether optically thick one-component models
of type A, B, or C (see Sect. 3.1 - 3.3) are at least able to
reproduce the NIR observations (ignoring the long-wavelength data).

\renewcommand{\RunNameLong}{1comp}
\renewcommand{\RunNameShort}{A}
\renewcommand{\DatasetName}{2}
\renewcommand{\ModelClassDescription}{simple one-component disks}
\renewcommand{\BestChiSquare}{$4.01$}
\StandardTableAndFigure

\subsection{Model \ModelNameShort}
For this goal, we compared the models in the parameter scanning run A
with data set 2.
The parameter ranges and results 
are shown in \StandardTableAndFigureList.
The $\chi^2_\mathrm{2,red}$ of the best model {\ModelNameShort} is {\BestChiSquare}.
The high $\chi^2_\mathrm{2,red}$ value of model {\ModelNameShort} indicates
that it cannot reproduce the NIR SED and visibilities.

\renewcommand{\RunNameLong}{1comp_curved}
\renewcommand{\RunNameShort}{B}
\renewcommand{\DatasetName}{2}
\renewcommand{\ModelClassDescription}{one-component disks with curved rims}
\renewcommand{\BestChiSquare}{$4.18$}
\StandardTableAndFigure

\subsection{Model \ModelNameShort}
We compared the models in the parameter scanning run B
with data set 2.
The parameter ranges and results 
are shown in \StandardTableAndFigureList.
The $\chi^2_\mathrm{2,red}$ of the best model {\ModelNameShort} is {\BestChiSquare}.
The high $\chi^2_\mathrm{2,red}$ value of model {\ModelNameShort} indicates
that it cannot reproduce the NIR SED and visibilities.

\renewcommand{\RunNameLong}{1comp_shadowed}
\renewcommand{\RunNameShort}{C}
\renewcommand{\DatasetName}{2}
\renewcommand{\ModelClassDescription}{one-component disks with shadowed regions}
\renewcommand{\BestChiSquare}{$3.67$}
\StandardTableAndFigure

\subsection{Model \ModelNameShort}
We compared the models in the parameter scanning run C
with data set 2.
The parameter ranges and results 
are shown in \StandardTableAndFigureList.
The $\chi^2_\mathrm{2,red}$ of the best model {\ModelNameShort} is {\BestChiSquare}.
The high $\chi^2_\mathrm{2,red}$ value of model {\ModelNameShort} indicates
that it cannot reproduce the NIR SED and visibilities.

\renewcommand{\RunNameLong}{2comp_ThinNull}
\renewcommand{\RunNameShort}{F}
\renewcommand{\DatasetName}{2}
\renewcommand{\ModelClassDescription}{optically thin dust at sub-AU scale}
\renewcommand{\BestChiSquare}{2.33}
\StandardTableAndFigure

\subsection{Model \ModelNameShort}
The purpose of parameter scanning run {\RunNameShort}
is to test whether data set 2 can be reproduced
by an NIR-emitting optically thin dust model
with only an optically thin dust component located at sub-AU
distance from the star.
The parameter ranges and results 
are shown in \StandardTableAndFigureList.
The model E2 can approximately reproduce data set 2,
with $\chi^2_\mathrm{2,red}=\BestChiSquare$.
In run E, we fix the $W_\mathrm{ring}$ of the optically thin ring
to $50\%$.
This is justified by the degeneracy between $W_\mathrm{ring}$ and $R_\mathrm{in}$,
which is illustrated in the next section.
Similarly, there is a degeneracy between $h$ and $\tau$ (see Sect. B.6),
and we therefore fix  $h$ to $0.5$.

\renewcommand{\RunNameLong}{2comp_ThinNull_Rin_Wring}
\renewcommand{\RunNameShort}{FA}
\renewcommand{\DatasetName}{2}
\renewcommand{\ModelClassDescription}{optically thin dust at sub-AU scale}
\renewcommand{\BestChiSquare}{$2.27$}
\StandardTableAndFigure

\subsection{Model scanning \RunNameShort}
The purpose of parameter scanning run {\RunNameShort}
is to explore the effect of changing $W_\mathrm{ring}$ and $R_\mathrm{in}$.
The parameter ranges and results 
are shown in \StandardTableAndFigureList.
There is a degeneracy between $W_\mathrm{ring}$ and $R_\mathrm{in}$.

\renewcommand{\RunNameLong}{2comp_ThinNull_h_tau}
\renewcommand{\RunNameShort}{FB}
\renewcommand{\DatasetName}{2}
\renewcommand{\ModelClassDescription}{optically thin dust at sub-AU scale}
\renewcommand{\BestChiSquare}{$2.49$}
\StandardTableAndFigure

\subsection{Model scanning \RunNameShort}
The purpose of parameter scanning run {\RunNameShort}
is to explore the effect of changing $h$ and $\tau$.
The parameter ranges and results 
are shown in \StandardTableAndFigureList.
There is a degeneracy between $h$ and $\tau$.

\fi

\bibliography{draft}

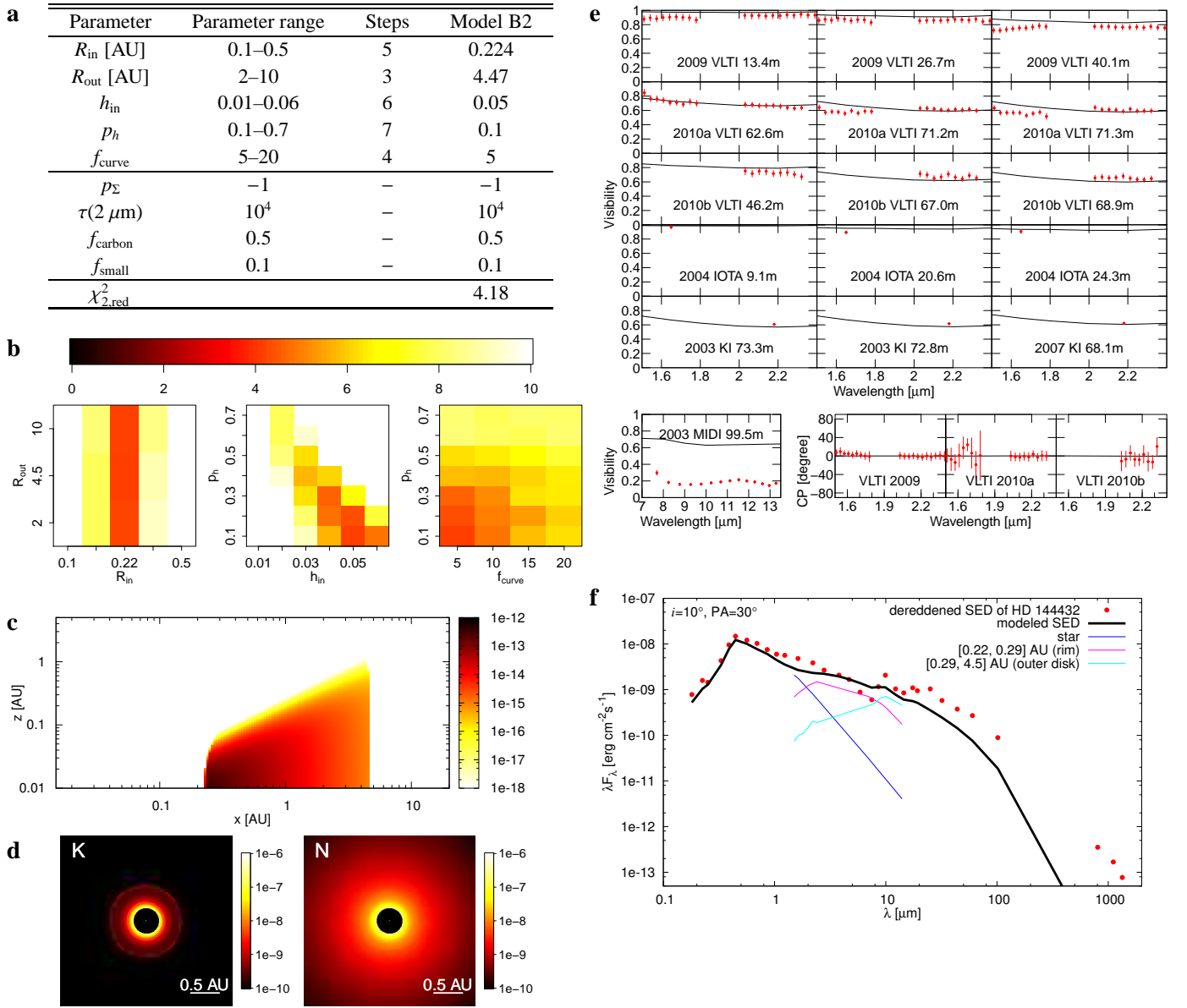
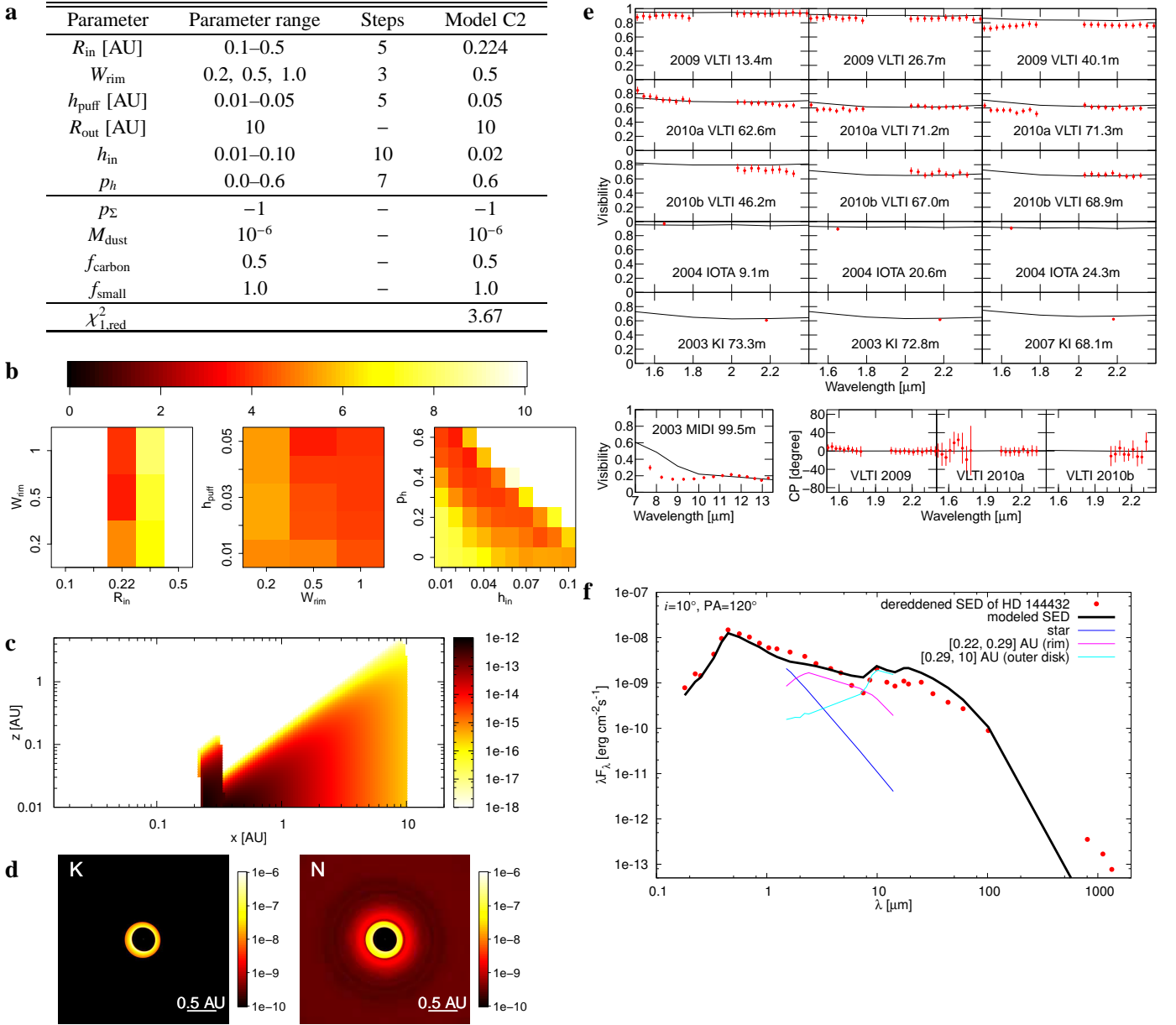
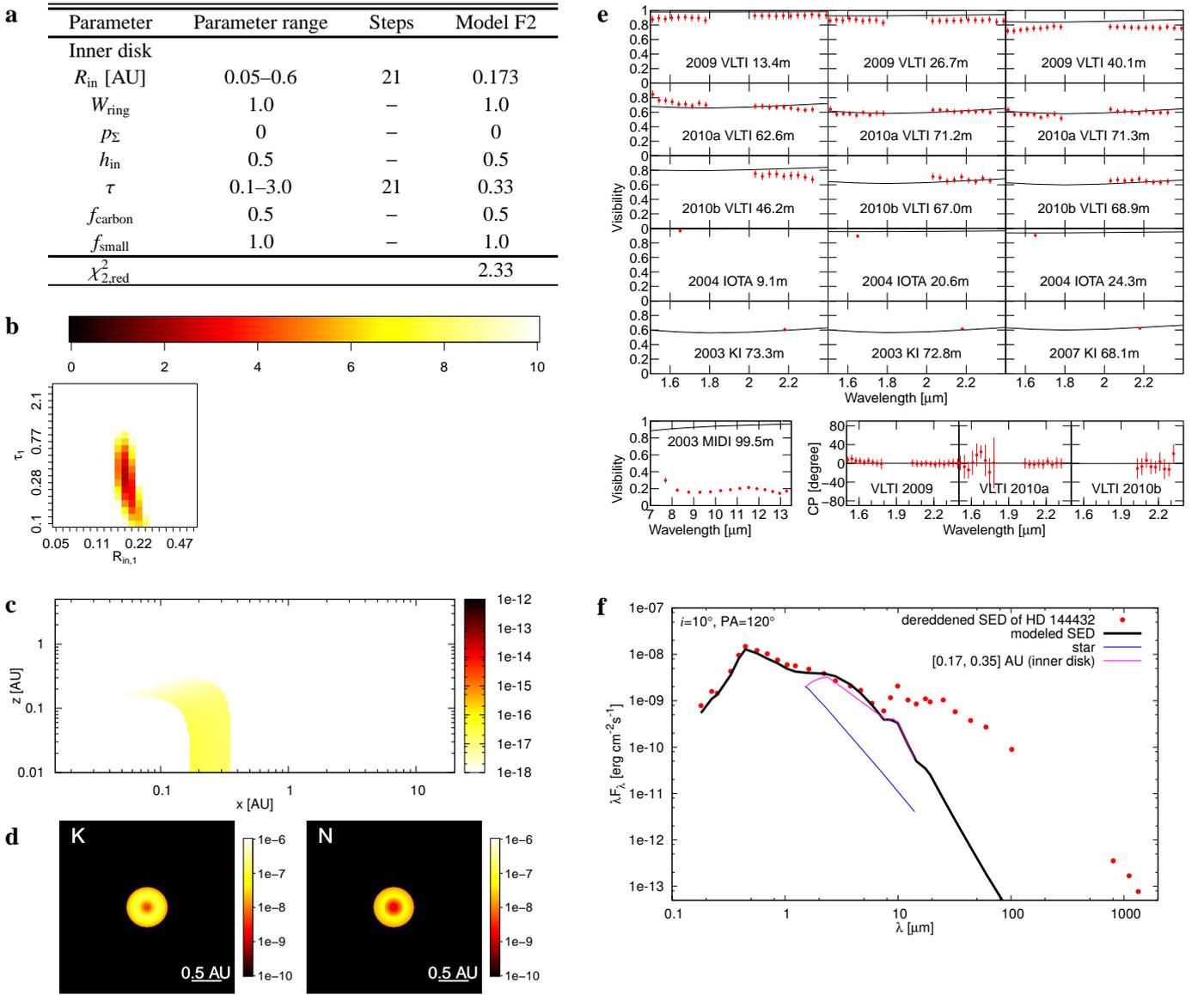
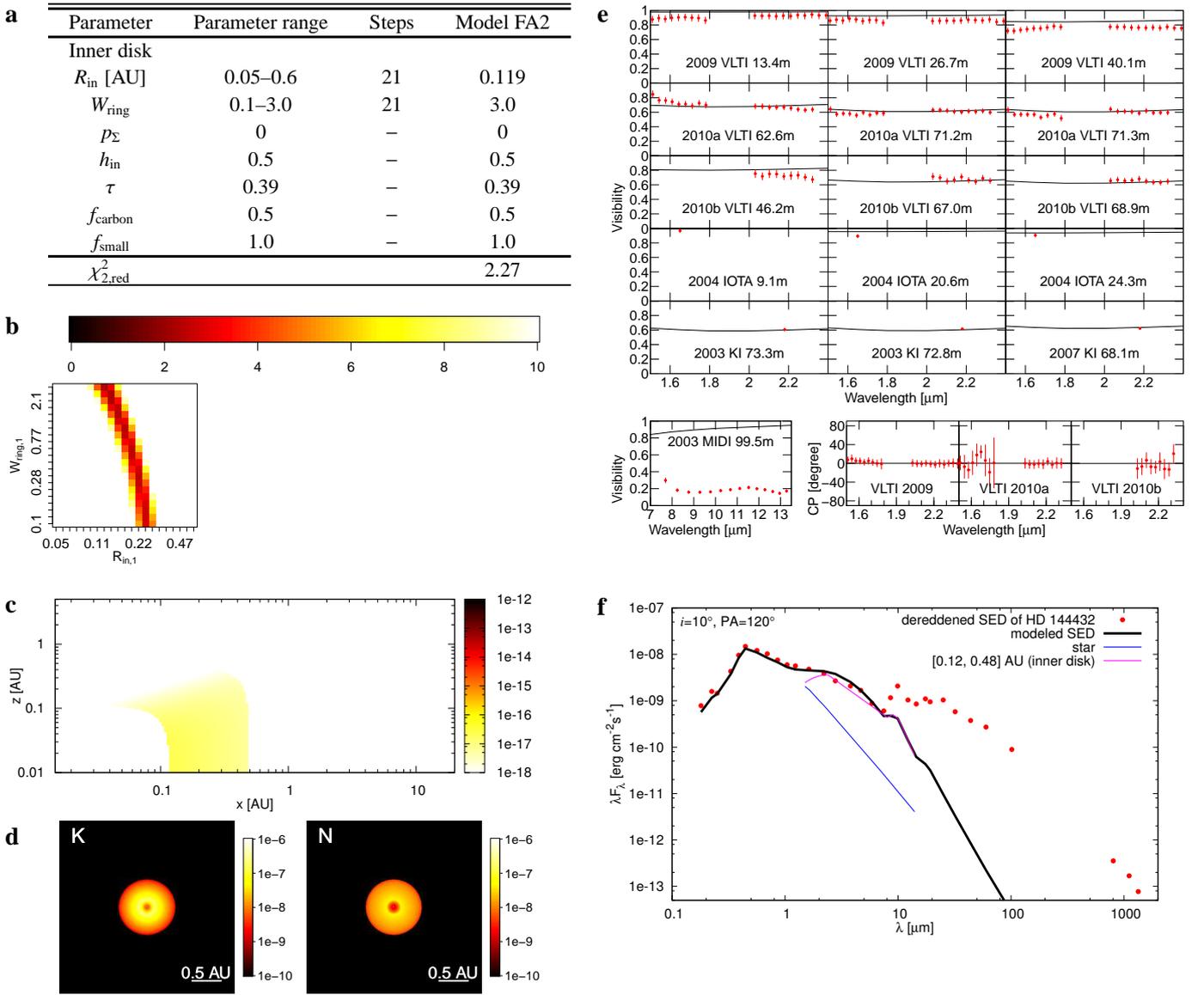
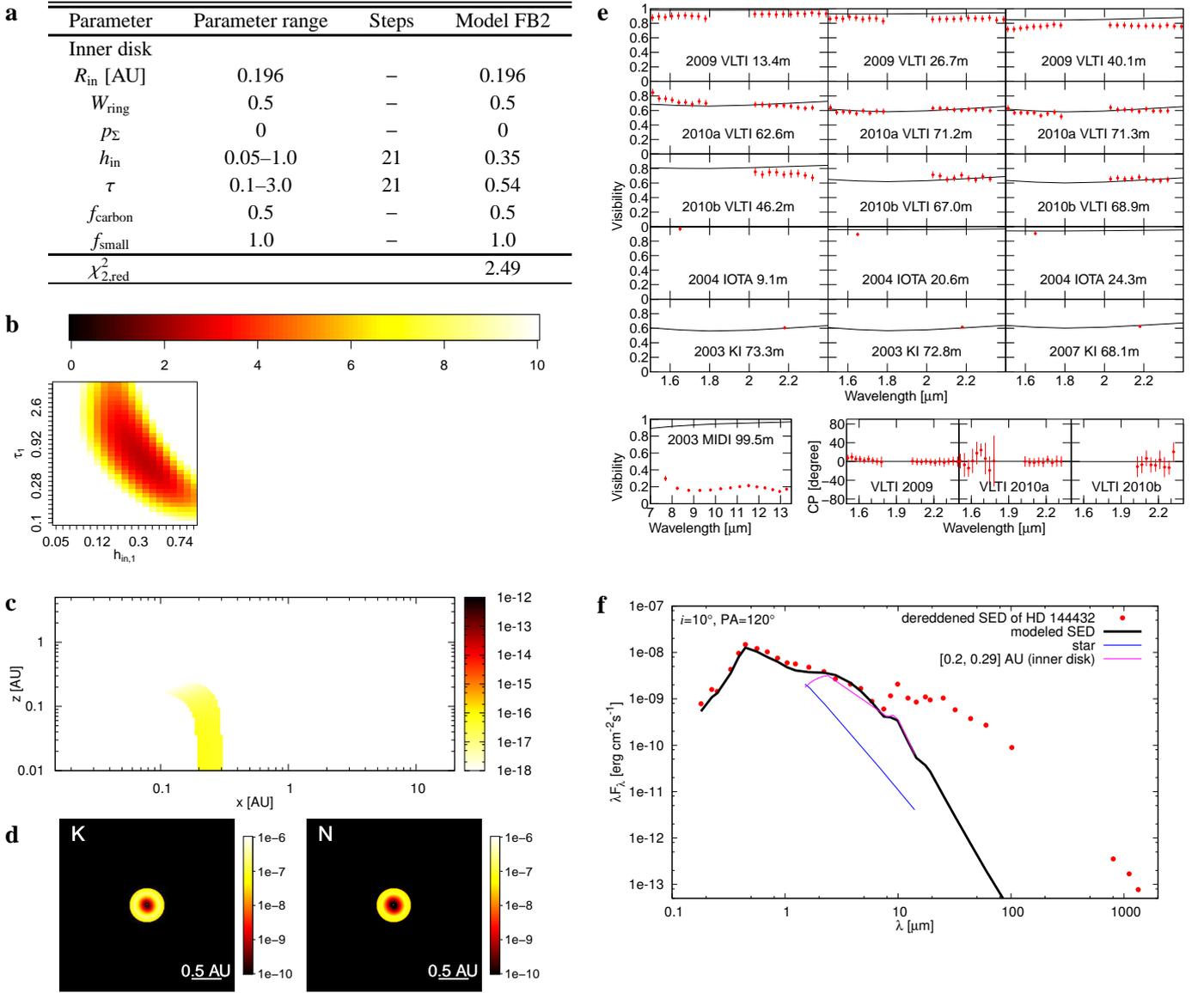
\end{document}